\documentclass[nofootinbib, reprint,
 preprintnumbers, amsmath, amssymb,
 aps, prd, superscriptaddress, floatfix]{revtex4-2}

\usepackage{mathtools}
\usepackage[dvipdfmx]{graphicx}
\usepackage{dcolumn}
\usepackage{bm}
\usepackage{physics}
\usepackage{hyperref}
\usepackage{braket}
\usepackage{float}

\usepackage{mathrsfs}

\usepackage{xcolor}
\usepackage{lipsum}

\begin{document}

\preprint{KEK-TH-2330}

\title{On the Feasibility of Bell Inequality Violation at ATLAS Experiment\\
with Flavor Entanglement of $B^0\bar{B}^0$ Pairs from $pp$ Collisions}

\author{Yosuke~Takubo}
\email{yosuke.takubo@kek.jp}
\affiliation{%
Institute of Particle and Nuclear Studies, High Energy Accelerator Research Organization (KEK), Ibaraki 305-0801, Japan.
}%
\affiliation{%
 The Graduate University for Advanced Studies (SOKENDAI), Hayama 240-0193, Japan.
}%

\author{Tsubasa~Ichikawa}
\email{ichikawa@qiqb.osaka-u.ac.jp}
\affiliation{%
Center for Quantum Information and Quantum Biology, Osaka University, Osaka 560-0043, Japan
}%

\author{Satoshi~Higashino}
\email{higashino@people.kobe-u.ac.jp}
\affiliation{%
 Department of Physics, Kobe University, Hyogo 657-8501, Japan.
}%

\author{Yuichiro~Mori}
\email{yuichiro@post.kek.jp}
\affiliation{%
Institute of Particle and Nuclear Studies, High Energy Accelerator Research Organization (KEK), Ibaraki 305-0801, Japan.
}%

\author{Kunihiro~Nagano}
\email{kunihiro.nagano@kek.jp}
\affiliation{%
Institute of Particle and Nuclear Studies, High Energy Accelerator Research Organization (KEK), Ibaraki 305-0801, Japan.
}%
\affiliation{%
 The Graduate University for Advanced Studies (SOKENDAI), Hayama 240-0193, Japan.
}%

\author{Izumi~Tsutsui}
\email{izumi.tsutsui@kek.jp}
\affiliation{%
 Institute of Particle and Nuclear Studies, High Energy Accelerator Research Organization (KEK), Ibaraki 305-0801, Japan.
 }%
\affiliation{%
 Department of Physics, The University of Tokyo, Tokyo 113-0033, Japan.
}%

\date{\today}

\begin{abstract}

We examine the feasibility of the Bell test ({\it i.e.}, detecting a violation of the Bell inequality) 
with the ATLAS detector in Large Hadron Collider (LHC) at CERN through the flavor entanglement between the $B$ mesons.
After addressing the possible issues that arise associated with the experiment and how they may be treated based on an analogy with conventional Bell tests, we show in our simulation study that under realistic conditions (expected from the LHC Run 3 operation) the Bell test is feasible under mild assumptions.  The definitive factor for this promising result lies primarily in the fact that the ATLAS detector is capable of measuring the decay times of the $B$ mesons independently, which was not available in the previous experiment with the Belle detector at KEK.   This result suggests the possibility of the Bell test in much higher energy domains and may open up a new arena for experimental studies of quantum foundations.


\end{abstract}
\maketitle

\section{INTRODUCTION}

Entanglement, or non-separability of quantum states, is {\it the} characteristic trait of quantum mechanics (QM), according to Schr\"odinger \cite{Schroedinger:1935}.
As pointed out in the seminal paper by Einstein, Podolsky, and Rosen (EPR) \cite{EPR35}, 
entanglement leads to mutual dependence of the (possible) measurement outcomes of local observables with respect to the constituents. This non-classical correlation inherent in the entanglement is what EPR employed to suggest that QM is incomplete as a physical theory, but the same correlation is now seen as a key resource to perform useful information processing in quantum information science \cite{Chitambar:2019}.

Entanglement is experimentally confirmed by detecting a violation of the Bell inequality \cite{Bell:1964, Goldstein:2011} which holds as long as the nature is governed by local realism, a principle deeply rooted in our scientific thought -- until the advent of QM at least.  As such, the 
test of the Bell inequality, or the Bell test in short, in actual physical phenomena has been an important subject of research in fundamental physics since the inequality was reformulated in form amenable to experiments \cite{CHSH:1969}.  In fact, many experiments have been conducted in the last four decades and reported violations in various systems: photons \cite{Aspect:1982, Tittel:1998, Weihs:1998, Giustina:2013, Giustina:2015, Shalm:2015, Larsson:2014, Rauch:2018}, ions \cite{Rowe:2001}, nucleons \cite{Sakai:2006}, superconducting phase qubits \cite{Ansmann:2009}, spin systems \cite{Hensen:2015, Dehollain:2016, Rosenfeld:2017}, and $B$ mesons observed in the Belle experiment in KEK \cite{Go:2004}, to name a few.

Among these tests, the test using the $B$ meson pairs \cite{Go:2004} is distinguished in that it is performed in a high energy experiment and invoked arguments \cite{Bertlmann:2004cr, Ichikawa:2008eq} on its validity. The theoretical basis of the Bell test using the $B$ meson pairs is laid on a formal analogy between the decay times of the $B$ mesons and the measurement angles for the spin measurement of the entangled spin pairs \cite{Gisin:2001}.  The experimental condition and the analogy jointly raise three issues to be addressed: 
First, in usual particle physics experiments, there is no room for a free choice in the decay times \cite{Bertlmann:2004cr} as required for the Bell test.
Second, since the $B$ mesons are unstable, the correlation between the flavors of the $B$ meson pair tends to be too weak to demonstrate the violation of the Bell inequality, unless some renormalization procedure \cite{Gisin:2001} is employed for the correlation function \cite{Bertlmann:2004cr}. 
Third, the Belle detector measures only the difference between the decay times of the $B$ mesons. This changes the upper bound for the Bell inequality, and makes the experimental confirmation of the violation extremely difficult \cite{Ichikawa:2008eq}.

In this paper, we examine the feasibility of the Bell test with the ATLAS experiment in Large Hadron Collider (LHC) at CERN through the flavor entanglement between the $B$ mesons.
We shall argue that the first issue may be dealt with by regarding the decay times as none other than a form of free choice,  since they are
generated purely stochastically \cite{Schmidt1970,Silverman2000,Walker2001} and hence can be treated similarly as random parameters utilized for conventional Bell tests ({\it e.g.}, \cite{Weihs:1998, Rauch:2018}).   This stochastic property also allows us to introduce the necessary renormalization procedure to overcome the second issue.  
Above all, the fact that the ATLAS detector is capable of measuring the decay times of the $B$ mesons independently gives a decisive advantage over the previous Bell test by using the Belle detector, solving the third issue almost thoroughly.  In brief, 
with analyses on experimental loopholes \cite{Clauser:1978, Garg:1987, Larsson:1998} and simulation study under realistic conditions during the LHC Run 3 operation from 2022 to 2026, we conclude that the Bell test is feasible by using the ATLAS detector on suitable assumptions. 
Our result suggests that the ATLAS detector enables the entanglement detection in an energy scale even higher than that of the Belle detector.

The rest of this paper is organized as follows. 
In Sec.~\ref{theory}, we provide a theoretical analysis on the feasibility of the flavor entanglement detection with the ATLAS detector. 
In Sec.~\ref{simulation}, after a detailed description of the flavor measurement and the identification process, we present simulation results which suggest that the $B$ mesons detected by the ATLAS detector are mutually spacelike separated and, accordingly, the detection of the violation of the Bell inequality should be possible. 
Section~\ref{conc} is devoted to our conclusion and discussions. 
For supporting our analysis, in the Appendix, we furnish another derivation of the Bell inequality without relying on the analogy mentioned above.

\section{Theoretical Basis}
\label{theory}
In this section, we develop a formulation to detect the entanglement by using the Bell inequality with the ATLAS detector. For deeper understanding of the outcomes given in
Sec.~\ref{simulation}, we begin with a brief summary of physics of $B$ meson pairs.
Next we introduce a modern
approach to derive the Bell inequality \cite{Bell:1964}. For detail,
see \cite{Goldstein:2011}. We mention the formal analogy between the decay times and the spin measurements in the usual Bell test, and show that the Bell inequality \cite{Ichikawa:2008eq} suitable to the flavor measurements of $B^0\bar{B}^0$ in the ATLAS experiment takes the same form as the standard Bell inequality.
We further show quantum violation of the Bell inequality, and finish
this section with discussions on experimental evaluation of the Bell inequality
and loopholes thereof. Note that the formulation given in this section
can be applied to other meson pairs such as $K^0{\bar K}^0$.
We hereafter work with the natural unit $\hbar=c=1$
when no confusion arises.

\subsection{Flavor measurements}
\label{setups}

The system we deal with is a pair of neutral $B$ mesons generated in the flavor singlet state,
\begin{equation}
|\psi\rangle=\frac{1}{\sqrt{2}}(|B^0\rangle|\bar{B}^0\rangle-|\bar{B}^0\rangle|B^0\rangle),
\label{psi}
\end{equation}
from $pp \to b\bar{b}$ processes. 
Here $\ket{B^0}$ and $\ket{\bar{B}^0}$ are the flavor eigenstates of the $B$ meson, which together form 
a complete orthonormal basis (flavor eigenbasis) in the two dimensional complex Hilbert space $\mathbb{C}^2$
describing the flavor internal degrees of freedom of a single meson.  
The total state space of the pair of $B$ mesons is given by the tensor product $\mathbb{C}^{2} \otimes \mathbb{C}^{2}$ of the respective  Hilbert spaces.

The state $|\psi\rangle$ is {\it entangled} as it cannot be factorized as a direct product of the one-particle states and, as such, it exhibits strong correlation between its constituents. One may thus expect that the correlation between the
neutral $B$ meson in $|\psi\rangle$ could trigger the violation of the Bell inequality under suitable measurement setups.

On the assumption of the unbroken $CP$ symmetry, the time evolution of a neutral $B$ meson obeys the Schr\"odinger equation
$
i\frac{d}{dt}|\psi\rangle=\hat{H}|\psi\rangle
$
with the phenomenological Hamiltonian \cite{BURAS1984369},
\begin{equation}
\hat{H}=
\left(\begin{array}{cc}
M-\frac{i}{2}\Gamma & M_{12}-\frac{i}{2}\Gamma_{12} \\
M_{12}-\frac{i}{2}\Gamma_{12} & M-\frac{i}{2}\Gamma
\end{array}\right)
\end{equation}
written in the flavor eigenbasis 
$
|B^0\rangle=
(
\begin{smallmatrix}
1\\
0
\end{smallmatrix}
)
$
and
$
|\bar{B}^0\rangle=
(
\begin{smallmatrix}
0\\
1
\end{smallmatrix}
)
$. 
On account of the symmetry,  the Hamiltonian $\hat{H}$ is invariant $(\mathcal{CP})\hat{H}(\mathcal{CP})=\hat{H}$ under the $CP$ transformation expressed as $|\bar{B}^{0}\rangle = \mathcal{CP} |B^{0}\rangle$ and $|B^{0}\rangle = \mathcal{CP} |\bar{B}^{0}\rangle$ in our convention.  

Note that $\langle B^0|\hat{H}|\bar{B}^0\rangle\neq0$ implies that the time evolution induces the flavor transition called the flavor mixing. Besides, the non-Hermitian Hamiltonian $\hat{H}$ describes the decay of the $B$ meson into other particles, resulting in the gradual decrease in the probability of remaining as the $B$ meson. 

Let us derive the explicit expression of the joint probability function in the flavors. The eigenstates of the phenomenological Hamiltonian $\hat{H}$ called the mass eigenstates take the forms,
\begin{eqnarray}
|B_{\rm H}\rangle=\frac{|B^0\rangle + |\bar{B}^0\rangle}{\sqrt{2}},
\qquad
|B_{\rm L}\rangle=\frac{|B^0\rangle - |\bar{B}^0\rangle}{\sqrt{2}},
\end{eqnarray}
with the eigenvalues,
\begin{eqnarray}
\lambda_{\rm H}=M_{\rm H}-\frac{i}{2}\Gamma_{\rm H},
\qquad
\lambda_{\rm L}=M_{\rm L}-\frac{i}{2}\Gamma_{\rm L},
\end{eqnarray}
where
\begin{eqnarray}
M_{\rm H}&=&M+M_{12}, 
\qquad
\Gamma_{\rm H}=\Gamma+\Gamma_{12}, \nonumber\\
M_{\rm L}&=&M-M_{12}, 
\qquad
\Gamma_{\rm L}=\Gamma-\Gamma_{12}.
\end{eqnarray}
Since the difference in the decay width is extremely small $\Gamma_{\rm H}-\Gamma_{\rm L}\approx0.001\Gamma$ \cite{pdg} and insignificant to the following discussions, we shall hereafter work on the approximation $\Gamma_{\rm H}=\Gamma_{\rm L}=\Gamma$ for simplicity.

The evolution of the mass eigenstates,
\begin{eqnarray}
|B_{\rm H}(t)\rangle=e^{-i\lambda_{\rm H}t}|B_{\rm H}\rangle,
\qquad
|B_{\rm L}(t)\rangle=e^{-i\lambda_{\rm L}t}|B_{\rm L}\rangle,
\end{eqnarray}
implies that the entangled state $|\psi\rangle$ generated at $t=0$ will become
\begin{eqnarray}
|\psi(t_1, t_2)\rangle=\frac{1}{\sqrt{2}}\left(|B_{\rm H}(t_1)\rangle|B_{\rm L}(t_2)\rangle-|B_{\rm L}(t_1)\rangle|B_{\rm H}(t_2)\rangle\right),\nonumber\\
\label{state_evo}
\end{eqnarray}
when the two mesons decay at the proper times $t_1$ and $t_2$, respectively.
By introducing dichotomic variables $A, B$ such that $A, B=+1$ for $B^0$ and $A, B=-1$ for $\bar{B}^0$, we obtain the quantum joint probability distribution of the flavor observations $A, B$ at the decay times $t_1, t_2$,
\begin{equation}
P^Q(A,B, t_1, t_2)=\frac{e^{-\Gamma(t_1+t_2)}}{4}\left(1-AB\cos(\Delta M\Delta t)\right),
\label{q_joint}
\end{equation}
with $\Delta t=t_1-t_2$ and $\Delta M=M_{\rm H}-M_{\rm L}=3.334\times 10^{-10}$~MeV \cite{pdg}  being the mass difference.
\subsection{The Bell inequality for flavor measurements}
\label{Bell_flavor}

Prior to discussing the Bell inequality for flavor measurements, let us briefly recall the Bell inequality itself, which 
was derived through Bell\rq s simple but careful analysis on the EPR argument \cite{EPR35} on QM.  In their argument,
EPR claimed that QM is incomplete as a physical theory on the basis of three assumptions: locality, our ability to freely choose experimental setups, and reality of physical quantities that should be dealt with a complete theory. Here, the reality means that the values of the physical quantities are determined or inferred with certainty, at least in principle.

Bearing the above historical background in mind, we shall introduce LRTs which fulfill
all the assumptions EPR made in their argument. 
Consider a measurement of spins of two spin $1/2$ particles, which are spacelike separated and have previously been interacted. Suppose that we are allowed to {\it freely choose} the experimental setups specified by parameters $a$ and $b$, respectively, for the two particles. More explicitly, we conduct the spin measurement for one particle along the measurement axis specified by $a$ and also for the other particle along the axis specified by $b$.  We then assign the values $A = +1$ if the outcome is along the axis measured (up spin) and $A = -1$ if it is opposite (down spin), and do the analogous assignment for $B$ for the second particle.   
Performing the above process of measurement many times, we obtain the probability distribution
$P_{a,b}(A,B)$
of finding the outcomes $A$, $B$ under the measurement setups $a$, $b$.

In the LRT description of the above experiment, associated with the {\it reality} of the physical quantities, we first suppose that there exist
parameters called hidden variables, collectively denoted by $\lambda$, which completely specify the states
of the physical systems with certainty. 
Making it explicit that the probability distribution $P_{a,b}(A,B)$ depends on $\lambda$, we write it as 
\begin{equation}
P_{a,b}(A,B)=\int P_{a,b}(A,B \, | \, \lambda)P(\lambda)\, d\lambda,
\label{prob_lambda}
\end{equation}
where $P_{a,b}(A,B \, | \, \lambda)$ is a conditional probability distribution of the outcomes $A$, $B$, given $\lambda$, and $P(\lambda)$ gives a distribution of $\lambda$ under the situation of measurement.
Note that LRTs are also called local hidden variable theories in literature.

The {\it locality} assumption implies that the choice of the measurement setup $a$ and the outcome $A$ are independent of that of $b$ and $B$, and vice versa. This allows us to decompose the conditional probability $P_{a,b}(A,B \, | \, \lambda)$ as
\begin{equation}
P_{a,b}(A,B\, | \, \lambda)=P_{a}(A \, | \, \lambda)P_{b}(B \, | \, \lambda),
\label{prob_ind}
\end{equation}
where $P_{a}(A \, | \, \lambda)$ and $P_{b}(B \, | \, \lambda)$ are the conditional probability distribution of the outcome $A$, $B$ with the measurement setups $a$, $b$ under the given $\lambda$.

We are now ready to introduce the Bell inequality. 
From Eq.~(\ref{prob_lambda}) and Eq.~(\ref{prob_ind}), the correlation between the outcomes $A$ and $B$ under the setups $a$ and $b$ reads
\begin{eqnarray}
C(a, b)
&=&\sum_{A,B}AB\,P_{a,b}(A,B) \nonumber\\
&=&\int A(a, \lambda)B(b, \lambda)P(\lambda)\, d\lambda,
\label{lrtcor}
\end{eqnarray}
where we have introduced
\begin{eqnarray}
A(a, \lambda)&=&P_{a}(A=1\, | \, \lambda)-P_{a}(A=-1\, | \, \lambda),\nonumber\\
B(b, \lambda)&=&P_{b}(B=1\, | \, \lambda)-P_{b}(B=-1\, | \, \lambda).
\end{eqnarray}
From $\vert A(a, \lambda) \vert \le1$ and $\vert B(b, \lambda)\vert\le1$, it is now straightforward to derive the Bell inequality
\begin{eqnarray}
|S|\le2
\label{Bell}
\end{eqnarray}
satisfied by the combination of four correlations
\begin{eqnarray}
S = C(a,b)+C(a^\prime,b)+C(a,b^\prime)-C(a^\prime,b^\prime)
\label{Bellcomb}
\end{eqnarray}
for any $a,a^\prime,b$ and $b^\prime$ (see, {\it e.g.}, \cite{Goldstein:2011}).

Many experiments have confirmed the violation of the Bell inequality \cite{Aspect:1982, Tittel:1998, Weihs:1998, Rowe:2001, Sakai:2006, Ansmann:2009, Giustina:2013, Larsson:2014, Giustina:2015, Shalm:2015, Hensen:2015, Dehollain:2016, Rosenfeld:2017, Rauch:2018}:
the experimental value of the LHS of Eq.~(\ref{Bell}) exceeds 2 with suitable
experimental setups and state preparation. In addition, these experimental
results are in good agreement with their quantum mechanical descriptions.
These observations clearly show that Nature, and its quantum mechanical description, do not satisfy at least one of the assumptions EPR made, even though they look apparently natural to hold.  

Now, to accomplish the task of putting the argument of the Bell inequality into the context of our flavor measurements, 
we first describe the measurement process of the flavor measurements and examine the analogy with the conventional argument 
leading to the Bell inequality mentioned above.  

In the flavor measurements of a pair of $B$ mesons, one meson decays at the proper time $t_1$ and its flavor, denoted by $A$, is revealed as either $+1$ for $B^0$ or $-1$ for $\bar{B}^0$. The other meson decays at the proper time $t_2$, and its flavor $B$ is $+1$ for $B^0$ and $-1$ for ${\bar B}^0$.  The outcomes of the measurements are then used to obtain the the joint probability distribution $P(A, B, t_1, t_2)$.  
The point here is that the proper times of the decays $t_1, t_2$ are determined stochastically. Thus, given the decay times $t_1, t_2$, the statistics of the flavor measurement is characterized by the conditional probability distribution $P_{t_1, t_2}(A,B)$, which 
is related to the joint probability distribution by
\begin{equation}
P_{t_1, t_2}(A,B)=\frac{P(A, B, t_1, t_2)}{\sum_{A,B}P(A, B, t_1, t_2)}.
\label{cond_form}
\end{equation}

At this point, one notices 
the apparent analogy between the decay times $t_1, t_2$ and the measurement parameters $a, b$, since the latter are also 
conditioning the probability distributions obtained under the setup specified by the parameters.  Assuming, for the moment, that 
this analogy holds perfectly, we realize that all the argument we just have gone through for the Bell inequality applies here as well.
It thus follows that, if we just formally replace $a, b$ with $t_1, t_2$, we end up with the same Bell inequality \eqref{Bell}
for the set of correlations \eqref{Bellcomb} where now we use
\begin{equation}
C(t_1, t_2)=\sum_{A,B}AB\, P_{t_1, t_2}(A,B)
\label{fl_cor}
\end{equation}
instead of $C(a, b)$ and the like.

We have seen before that in QM the joint probability distribution $P^Q(A, B, t_1, t_2)$ is given by Eq.~(\ref{q_joint}).  
From the relation \eqref{cond_form}, one then finds the corresponding conditional probability distribution,
\begin{equation}
P_{t_1, t_2}^Q(A,B)=\frac{1}{4}(1-AB\cos(\Delta M\Delta t)),
\end{equation}
and also from Eq.~\eqref{fl_cor} the quantum correlation,
\begin{equation}
C^Q(t_1, t_2)=\sum_{A,B}AB\, P^Q_{t_1, t_2}(A,B)=-\cos(\Delta M\Delta t),
\label{q_correl}
\end{equation}
which is the renormalized correlation function introduced in \cite{Gisin:2001}, 
and free from the exponential decay law that the joint probability distribution (\ref{q_joint}) suffers.
Note that $C^Q(t,t)=-1$ implies perfect anti-correlation in the flavors of the $B$-meson pair decaying at the same proper time $t$.

To proceed, let us consider the special case of the decay times,
\begin{equation}
t_2-t_1^\prime=t_1-t_2=t_2^\prime-t_1=\Delta t,
\label{time_cond}
\end{equation}
which corresponds to a typical configuration of the measurement setups for the experimental verification of the Bell inequality.
Indeed, if we denote by $S^Q(\Delta t)$ the combination $S$ in Eq.~\eqref{Bellcomb} when we use the quantum correlation function (\ref{q_correl}) for
Eq.~\eqref{time_cond}, we obtain 
\begin{eqnarray}
&&S^Q(\Delta t)\nonumber\\ 
&&=C^Q(t_1,t_2)+C^Q(t_1^\prime,t_2)+C^Q(t_1,t_2^\prime)-C^Q(t_1^\prime,t_2^\prime)\nonumber\\
&&= -3\cos(\Delta M\Delta t) + \cos(3\Delta M\Delta t).
\label{q_s}
\end{eqnarray}
By differentiating $S^Q(\Delta t)$ with respect to $\Delta t$, we easily find that $|S^Q(\Delta t)|\le 2\sqrt{2}$ and
the maximum value $2\sqrt{2}$ is attained at $\Delta t=\pi/4\Delta M\approx1.55$ ps.  This indicates that 
we may observe violation of the Bell inequality with a pair of $B$ mesons as well, once the measurement is carried out properly.  

For comparison, as an example of LRT models we mention the spontaneous disentanglement model \cite{Go:2007ww}, where
one obtains the conditional correlation function,
\begin{equation}
C^S(t_1, t_2)=-\cos(\Delta Mt_1)\cos(\Delta Mt_2).
\label{SD}
\end{equation}
Observe that the form \eqref{SD} fits in the formula \eqref{lrtcor} of LRT (with $a, b$ replaced by $t_1, t_2$) if we let $A(t_1, \lambda)=-\cos(\Delta Mt_1)$ and $B(t_2, \lambda)=\cos(\Delta Mt_2)$ under the use of the normalization condition $\int P(\lambda)\, d\lambda=1$.  It thus follows that the correlation (\ref{SD}) obeys the Bell inequality trivially.

Now, coming back to the question of the validity of analogy between the parameters $a, b$ and $t_1, t_2$, it has been argued \cite{Bertlmann:2004cr} that, while the former can be chosen at will by the experimenter, the latter are determined by nature and cannot be altered freely.  Although this is apparently the case in reality, one may take the viewpoint that the experimenter should also be influenced by nature and hence, logically speaking, one cannot deny completely the possibility of the parameters $a, b$ being determined by other sources including the hidden variables $\lambda$.  To avoid impractical impasse, it is customary to accept the choice of $a, b$ performed by some random number generator (RNG) as a result of free will.  By the same token, one may accept the choice of $t_1, t_2$ performed by the particles in their random decays as an act of free will, given that such random decays have actually been utilized as a source of quantum RNG 
\cite{Schmidt1970,Silverman2000,Walker2001,Giustina:2015,Hensen:2015,Shalm:2015}.  

To elaborate this idea a little more, we recall the fact that in our experiment the target system of measurement is the flavor subspace which is a part of the entire space of freedoms possessed by the $B$ meson.  On the other hand, the decay times of the $B$ meson are governed and determined quantum mechanically by a separate part of the system,  which may be regarded as a quantum RNG equipped with the particle working independently from the flavor part.  This picture will then allow us to put our experiment on a par with preceding Bell tests as far as the free will (or freedom-of-choice) loophole is concerned.  
This viewpoint has apparently been adopted in the earlier analysis of the Belle experiment \cite{Go:2007ww}, but we shall also mention in the Appendix an alternative argument to retain the formal structure of the Bell inequality referring to earlier works \cite{Ichikawa:2008eq}.

\subsection{Note on experiments and loopholes}

The ATLAS experiment can measure the decay times $t_1, t_2$ of the $B$ meson pair (\ref{state_evo}) independently, which enables us to evaluate the correlation according to Eq.~\eqref{fl_cor} and thereby obtain the Bell inequality $|S(\Delta t)|\le 2$ for the combination \eqref{time_cond}.  This is a crucial advantage over the Belle experiment where we measure the events only through their difference $\Delta t$ in the decay times, resulting in the increase in the upper bound of the Bell inequality, making the Bell test difficult accordingly \cite{Ichikawa:2008eq}.

Given this prospect, we now wish to address, in addition to the free will loophole we have just mentioned, 
two other major loopholes \cite{Giustina:2015,Hensen:2015,Shalm:2015} that may hamper the Bell test with the ATLAS experiment.
One is the efficiency loophole, which concerns that a certain proportion of unobserved events may enable LRT to exceed the upper bound of the Bell inequality (\ref{Bell}), which can be excluded only if the detection efficiency is greater than $2\sqrt{2}-2\approx82.8\%$ \cite{Garg:1987, Larsson:1998}.  If not, we are basically forced to make the fair sampling assumption \cite{Clauser:1978} that the detection probability is independent of the measurement setups $a, b$.  The combination of the correlations \eqref{Bellcomb} evaluated from the actually observed events is then assured to be identical with that evaluated from the total events including unobserved ones, and this ensures that the experimental violation of the Bell inequality implies incompatibility of the assumptions EPR made.  Unfortunately, with the ATLAS experiment, the detection 
efficiency is only 2.0\% as shown in Sec.~\ref{selection} due to the loss in event selection 
processes. This implies that we need to make the fair sampling assumption that the probability of the detection of the decay is independent of the decay times $t_1, t_2$, which looks fairly reasonable and has certainly been the case in usual measurements of decay times.

\begin{figure}[tb]
\begin{center}
\includegraphics[width=8cm]{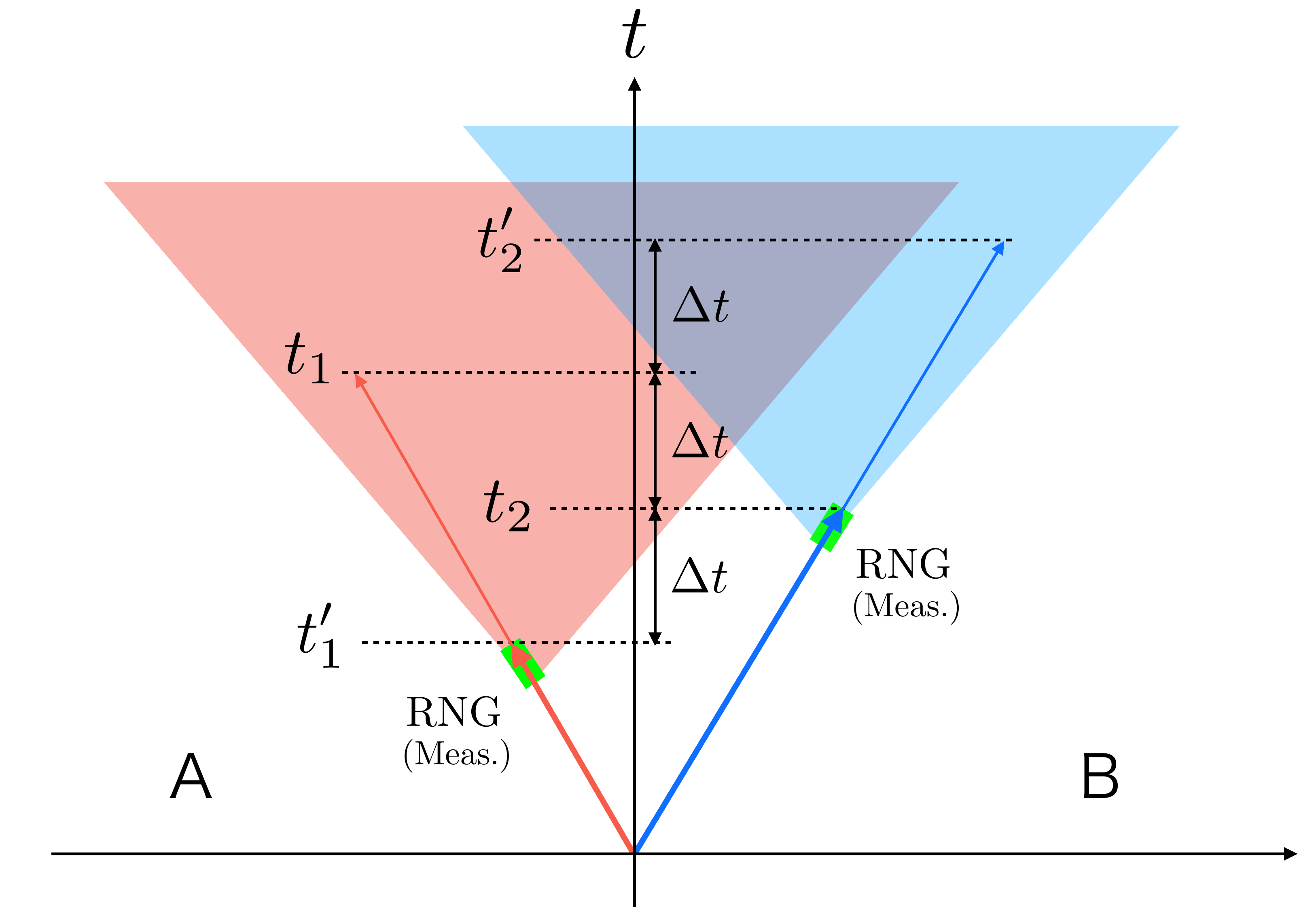}
\caption{(Color online) Minkowski diagrams for the spacetime events related to the decay of a pair of $B$ mesons considered for the Bell test.  After the $pp$ collision emerge a pair of $B$ mesons, which subsequently decay at $t_2$ (or $t'_2$) in the region (A) and at $t_1$ (or $t'_1$) in the region (B).  The red and blue solid lines depict the actual spacetime trajectories (world lines) of the respective $B$ mesons, while the shaded zones depict the forward light cones of the activated times of the RNG embedded in the $B$ mesons in the first two decays.  The flavor measurements, which are to be performed simultaneously with the RNG within a typical short period of time allocated for weak interactions colored in green, are completed retroactively after the decay modes are determined by identifying the decayed particles.
For the combination of the correlations $S(\Delta t)$ used for the Bell test, the locality condition requires that the final measurement at $t'_2$ be completed before the information of the first measurement at $t'_1$ reaches.  Our simulation indicates that this condition can be fulfilled with the ATLAS experiment.  
}
\label{fig:locality}
\end{center}
\end{figure}

The other loophole is the locality loophole \cite{Clauser:1978}, which concerns whether
the experimental setups guarantee the locality assumption: if the actual measurement configuration allows the measurement setup $a$ or outcome $A$ to affect $b$ or $B$, the locality assumption no longer holds, invalidating the direct link between the experimental violation of the Bell inequality and the incompatibility of the assumptions.  On account of the RNG embedded in the $B$ meson, and also due to the purely quantum nature of the decay, the operating time of the RNG is interpreted as the duration of the decay itself, which is of the order assigned to typical weak interactions.  Besides, as we shall see in Sec.~\ref{sec:ana_results}, it is possible to select the decay events occurring in the ATLAS detector so that the pair is mostly spacelike separated (see Fig.\ref{fig:locality}).  This indicates that the locality loophole can be closed virtually with the ATLAS experiment.

\section{FEASIBILITY STUDY}
\label{simulation}
\subsection{Outline}
The ATLAS experiment is performed at the CERN LHC in order to study phenomena in proton-proton ($pp$) and heavy-ion collisions. The ATLAS detector \cite{Aad:2008zzm}, which is designed for general physics purposes, consists of a superconducting solenoid surrounding the inner detector (ID) and a large superconducting toroid magnet system with muon detectors enclosing the electromagnetic and hadron calorimeters. 

The ATLAS experiment collected 5.1~fb$^{-1}$ of $pp$ collision data with the total center of mass energy $\sqrt{s}$~=~7~TeV and 21.3~fb$^{-1}$ with 8~TeV in Run~1 (2010-12), and 149~fb$^{-1}$ with 13~TeV in Run~2 (2015-2018). The instantaneous luminosity of $pp$ collisions has been increased during the operation and reached a maximal value of $2.2~\times~10^{34}$~cm$^{-2}$s$^{-1}$ in 2017, more than twice of the LHC design value ($1.0~\times~10^{34}$~cm$^{-2}$s$^{-1}$). The experiment called Run~3 is expected to start with 14~TeV of $pp$ colliding energy in 2022 after the long-shutdown~2 (2019-2021), and is expected to collect 180~fb$^{-1}$ of data by the end of 2026.

On the basis of the above specifications, we carried out a simulation study to evaluate the feasibility of the Bell test by means of the flavor entanglement of the $B$ meson pair in the ATLAS experiment. In this simulation, we assumed that nature obeys QM, and performed an error analysis of $S^{Q}(\Delta t)$ evaluated from the decay modes $B^{0} \to D^{\ast-} \mu^{+} \nu$ ($D^{\ast-} \to D^{0} \pi^{-}$, $D^{0} \to K^{+} \pi^{-}$) and their charge conjugate modes, which were previously employed in the flavor entanglement detection at the Belle experiment \cite{Bertlmann:2004cr, Go:2007ww}. In $B^{0} \to D^{\ast-} \mu^{+} \nu$ events, in practice, the flavor of a neutral $B$ meson at the decay can be identified from the configuration of the electric charges of the decay products, {\it i.e.}, $\mu^{+}\pi^{-}\pi^{-}K^{+}$ ($\mu^{-}\pi^{+}\pi^{+}K^{-}$) from $B^{0}$ ($\bar{B}^{0}$) decay.

Our feasibility study with our custom made simulation consists of three steps: (i) event generation, (ii) signal selection, and (iii) background estimation, each of which is presented in detail in the following three subsections, respectively. These three steps as a whole sift the events coming from the entangled state \eqref{psi}, and enable us to compute $S^{Q}(\Delta t)$. The simulation results are given in the last subsection.

\subsection{Event generation}
As the experimental condition to generate the events in our simulation, 14 TeV of $pp$ colliding energy was assumed. 
We used PYTHIA 8.245 \cite{pythia} to generate $gg/qq \to b\bar{b}$ in $pp$ collisions as well as $b\bar{b}$ pairs associated with di-jet events of light quarks. The production cross section of $B^{0}\bar{B}^{0}$ pairs is 69~$\mu$b in $gg$ and $qq$ interactions, and 319~$\mu$b in di-jet events of light quarks. 
$B^{0}$ generated in these processes is forced to decay into $D^{\ast-}\mu^{+}\nu$ ($D^{\ast -} \to D^{0} \pi^{-}$, $D^{0} \to K^{+} \pi^{-}$). The cross sections of such final states are $pp \to B^{0}\bar{B}^{0}$ (4.5~nb), $pp \to B^{\ast 0}\bar{B}^{\ast 0}$ (21.8~nb) and $pp \to B^{0}\bar{B}^{\ast0}$ (19.8~nb) in $gg$ and $qq$ interactions, respectively. 


As shown in Sec.~\ref{setups}, assuming that the initial state is the entangled state \eqref{psi} in QM, the flavors of  $B^{0}\bar{B}^{0}$ pair oscillate, and then those of the remaining $B$ mesons evolve with the joint probability distribution \eqref{q_joint}.
Then $|S^{Q}(\Delta t)|$ attains the maximal value $2\sqrt{2}$ at $\Delta t=1.55$~ps (See Sec.~\ref{Bell_flavor}).


Measurement of decay time of $B^{0}$ is crucial to measure $S^Q(\Delta t)$. In an ATLAS measurement \cite{Aaboud:2016bro}, the resolution of the proper decay position of  $B^{0}$ ($L^{B}_{\mathrm{prop}} = ct$) was estimated as 34~$\mu$m in $B^{0} \to J/\psi K_{S}$ and $B^{0} \to J/\psi K^{\ast0}$ decays, which are obtained from the vertex fit of the two muons from a $J/\psi$ decay. In $B^{0} \to D^{\ast-} \mu^{+} \nu$, $\mu^{+}$ and $\pi^{-}$ from $D^{\ast-}$ decay can be used for the vertex fit to measure $L^{B}_{\mathrm{prop}}$, and the resolution is expected to be similar. For that reason, 34~$\mu$m is assumed as $L^{B}_{\mathrm{prop}}$ resolution in this analysis and accordingly that of decay time ($L^{B}_{\mathrm{prop}}/c$) is 0.11~ps. 

In addition, the special LHC operation with the number of $pp$ collisions ($\mu$) around one per beam bunch crossing  (so-called low-$\mu$ run) and 1~fb$^{-1}$ of an integrated luminosity are assumed to suppress combinatorial backgrounds, which are expected to be the main source of backgrounds in this study. We describe the detail of the background estimation in Sec.~\ref{sec:background}. 

\subsection{Signal selection}
\label{selection}
We applied acceptance and selection cuts to the truth level information for two neutral $B$ mesons decaying into $D^{\ast \mp} \mu^{\pm} \nu$, by following the procedure used in measurement to evaluate the $b$-hadron production cross section from the decay modes to $D^{\ast+}\mu^{-}X$ final states in $pp$ collisions in the ATLAS experiment with $\sqrt{s} = 7$~TeV \cite{Aad:2012jga}. 

The overall selection efficiency $\epsilon$ is given as a product of the reconstruction efficiency $\epsilon_{\mathrm{reco}}$, muon trigger efficiency $\epsilon_{\mathrm{trigger}}$ and selection efficiency $\epsilon_{\mathrm{selection}}$. We set $\epsilon_{\mathrm{reco}} = 0.483$ by following the evaluations in \cite{Aad:2012jga}. In our analysis, di-muon trigger with 4~GeV of transverse momentum ($p_{\mathrm{T}}$) threshold is assumed, in contrast to 6~GeV in a single muon trigger used in \cite{Aad:2012jga}. Taking into account the trigger performance in the ATLAS experiment, we set $\epsilon_{\mathrm{trigger}} = 0.429$\footnote{The efficiency for the single muon trigger with $p_{\mathrm{T}} > 6$ GeV is evaluated as 0.819 in \cite{Aad:2012jga}. We assume that the efficiency drops by a factor of 0.8 (i.e., $0.819\times0.8\approx0.655$) by setting $p_{\mathrm{T}}$ threshold to 4~GeV, and then $\epsilon_{\mathrm{trigger}}$ becomes 0.429 ($\approx0.655^2$), requiring the criteria for two muons.}. 

For the event selection, $p_{\mathrm{T}}$ above 1~GeV was required to $\pi^{-}$ ($\pi^{+}$) and $K^{+}$ ($K^{-}$) from a $D^{0}$ ($\bar{D}^{0}$) decay as well as $p_{\mathrm{T}}$ above 250~MeV to  $\pi^{-}$ ($\pi^{+}$) from a $D^{\ast -}$ ($D^{\ast +}$) decay. In addition, taking into account that the invariant mass of the decay modes from two neutral $B$ mesons is within one sigma interval of the distributions 0.68, we assumed efficiency of 0.46 (=$0.68^{2}$) as the invariant mass cut for each of $K^{\mp}\pi^{\pm}$ and $D^{\ast \pm} \mu^{\mp}$.

The event selection cuts mentioned above results in $\epsilon_{\mathrm{selection}} = 0.097$. Multiplying $\epsilon_{\mathrm{reco}}$, $\epsilon_{\mathrm{trigger}}$ and $\epsilon_{\mathrm{selection}}$,  we obtained $\epsilon = 0.02$. 

As the acceptance cut, existence of two muons with $p_{\mathrm{T}}$ being above 4~GeV and pseudorapidity $|\eta|$ bellow 2.4 are required. In addition, requirement on two $D^{\ast \pm}$ mesons with $p_{\mathrm{T}} > 2.25$~GeV and $|\eta| < 2.5$ were applied. With these cuts, the acceptance $A$ and cross section times acceptance $\sigma \times A$ become $1.49 \times 10^{-3}$ and 369~pb, respectively, for the signal events.

Summing up, the number of the signal events left after the acceptance and selection cuts can be evaluated as $(\sigma \times A) \times \epsilon \times L = 7.4 L$, where $L$ is an integrated luminosity in unit of pb$^{-1}$. With this formula, by assuming $L=1\times10^3~{\rm pb}^{-1}$, we obtained the expected number of the signal events $7.4 \times 10^{3}$.

\subsection{Background estimation} \label{sec:background}
We have considered two possible sources of the background on our analysis.
First, two neutral $B$ mesons can be created from different gluons and they may contribute as irreducible background. 
Our simulation showed that such background is less than 0.1\% with respect to the signal and, therefore, negligible.

The other possible background is the combinatorial background that is caused by mis-reconstructed signals with particles from different origins. In measurements of decays to $D^{\ast+}\mu^{-}X$ final states in the ATLAS experiment \cite{Aad:2012jga}, the background contamination is $6.8 \pm 0.26$\%, where that of the combinatorial backgrounds is 6.2\%. One of such backgrounds is the mis-identification of the two decay modes $D^{\ast+}$ from $c \to D^{\ast+}X$ and $\mu^{-}$ from $\bar{c} \to \mu^{-}X'$, and those from $b \to D^{\ast+}\mu^{-}X$. 

In our analysis, we applied the cuts, which are the same as \cite{Aad:2012jga} except for the $p_{\mathrm{T}}$ thresholds of 4~GeV and 2.25~GeV instead of 6~GeV and 4.5~GeV for muon and $D^{\ast \pm}$, respectively. Our analysis assumes approximately one $pp$ interaction per beam bunch crossing, whereas that is above two for most of the period during which data for \cite{Aad:2012jga} were collected (between August and October 2010) \cite{atlas_mu}. In addition, the selection cuts are applied to two neutral $B$ mesons instead of one. For those reasons, the background contamination is expected to be smaller than 6.8\% in our measurement. Only the simulated signal events are used in our analysis, and the backgrounds are conservatively considered as systematic uncertainty as described in Sec.~\ref{sec:ana_results}.


\subsection{Analysis results} \label{sec:ana_results}
In our simulation, we examined two issues. One is whether the locality condition is satisfied with the events detected by the ATLAS detector. The other is how much the Bell inequality violates, with statistical and systematic errors considered.

\begin{figure}[tb]
\begin{center}
\includegraphics[width=8cm]{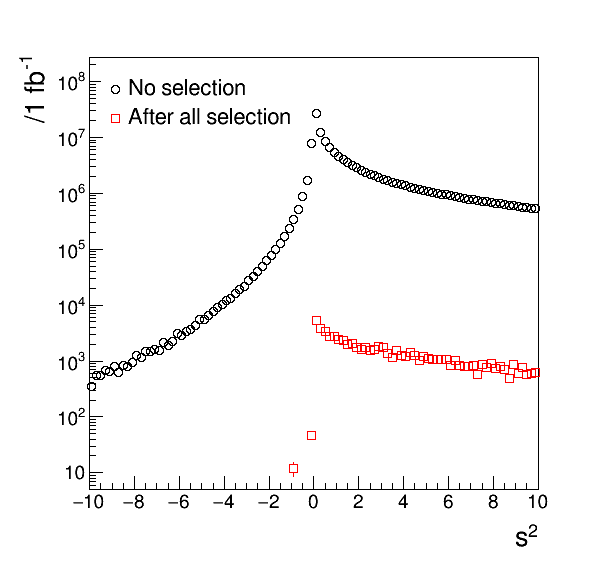}
\caption{(Color online) Distributions of the squared proper distances $s^2$ of the $B^0\bar{B}^0$ decay events before and after the acceptance and selection cuts. The events are spacelike when $s^2>0$.}
\label{fig:ss}
\end{center}
\end{figure}

Figure~\ref{fig:ss} shows the distributions of the squared proper distance $s^{2} = -c^{2}\Delta t^{2} + \Delta L^{2}$ of the $B^{0}\bar{B}^{0}$ decay events before and after the acceptance and selection cuts. Most of the events are spacelike ($s^{2} > 0$) even without any cuts, and more than 99\% events are spacelike after the cuts. The locality condition is thus perfectly satisfied in our analysis.

The quantum correlation $C^Q(t_1, t_2)$ can be calculated from the experimental data by using the following formula:
\begin{eqnarray}
C^Q(t_1, t_2) = \frac{\sum_{A,B}AB\, N^Q_{t_1,t_2}(A, B) }{\sum_{A,B}N^Q_{t_1,t_2}(A, B)}, 
\end{eqnarray}
where $N^Q_{t_1,t_2}(A, B)$ ($A,B$ are 1 for $B^0$ and $-1$ for $\bar{B}^0$) is the number of the events that two neutral $B$ mesons decay into the flavor $A$ at $t_1$ and $B$ at $t_2$, respectively. $S^{Q}(\Delta t)$ in Eq.~(\ref{q_s}) is calculated from the $C^Q(t_1, t_2)$ distribution under the configuration \eqref{time_cond}.

\begin{figure}[tb]
\begin{center}
\includegraphics[width=8cm]{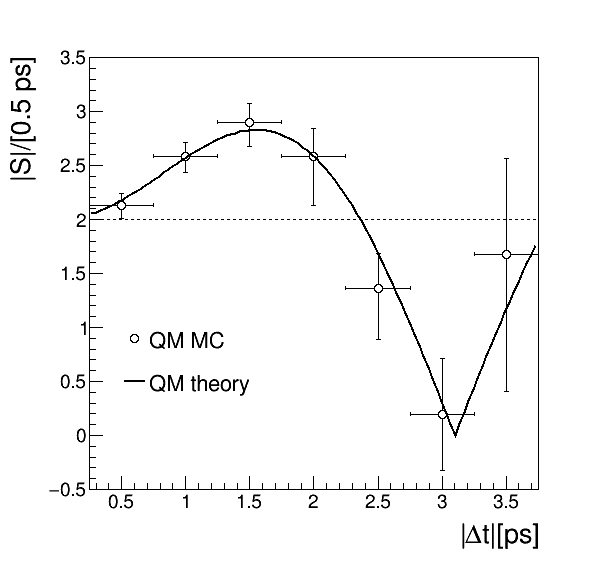}
\caption{$|S^Q(|\Delta t|)|$ after the acceptance and selection cuts in the case of QM with the theory line. The dotted line at $|S(|\Delta t|)| = 2$ shows the upper bound of the Bell inequality.}
\label{fig:s}
\end{center}
\end{figure}

Figure \ref{fig:s} shows $|S^{Q}(|\Delta t|)|$ after the acceptance and selection cuts in the case of QM. As the statistical error, only that of the signal is taken into account, since the contribution from the background is negligible as discussed in Sec. \ref{sec:background}. 

We evaluated the systematic error, by considering the worst case where the backgrounds contaminate only one $\Delta t$ bin in $C^{Q}(t_{1}, t_{2})$.
Since the combinatorial backgrounds are the main source and should contribute equally to $B^{0}$ and $\bar{B}^{0}$, 
it is assumed that the same fraction (50\%) of the backgrounds contaminates the same ($N^Q_{t_1,t_2}(A, B)$ for $B^0B^0$ or $\bar{B}^0\bar{B}^0$)  and opposite flavors of two neutral $B$ mesons ($N^Q_{t_1,t_2}(A, B)$ for $B^0\bar{B}^0$ or $\bar{B}^0B^0$). The amount of the backgrounds is assumed as 0.26\% which is assigned as the systematic error on the background contamination in \cite{Aad:2012jga}. This systematic error is the largest contribution to the shift from $|S^{Q}(|\Delta t|)|$ in this treatment. In addition, the systematic error originated from $\Delta t$ resolution, 0.16 ($0.11 \times \sqrt{2}$)~ps, is taken into account.


The significant excess from the upper bound of the Bell inequality \eqref{Bell} is obtained with $|S^{Q}(|\Delta t|)| = 2.89 \pm 0.17 (\mathrm{stat.}) ^{+0.06}_{-0.13} (\mathrm{syst.})$, where $\Delta t = 1.5 \pm 0.25~\mathrm{ps}$. In the systematic error, contribution from the backgrounds is (-0.11, +0) and that from $\Delta t$ resolution is $\pm 0.06$. 
This result demonstrates that the flavor entanglement detection using the Bell inequality --- or the Bell test with the flavor entanglement --- is feasible in the ATLAS experiment.

\section{CONCLUSION AND DISCUSSIONS}
\label{conc}

The Bell inequality is a principal touchstone of testing the local realism posited by Einstein at the time of the formation of quantum theory.  As the history shows, violations of the Bell inequality have been found time and time again, which reject
the local realism with the measured systems of 
photons, electrons or nucleons at low energies.  Extending to systems with higher energies will be important 
for establishing the nonlocal nature universally, and here we present a simulation study on the feasibility of the Bell test by means of flavor entanglement of a pair of $B$ mesons in the ATLAS experiment at CERN.  Our simulation resulted in the affirmative: we will find the maximal violation of the Bell inequality at the time difference $\Delta t\approx1.5$ ps in the decays of the two entangled $B$ mesons, rejecting yet again the local realism at the highest energy scale 14 TeV ever.  

This will be the first case of violation of the Bell inequality in the community of particle physics experiment, given that the earlier analysis with the Belle experiment \cite{Go:2004} was found to be inconclusive, due primarily to the lack of selection process of spacelike events and that of the independent identification of the decay times. The former leads to the locality loophole, whereas the latter results in the increase of the upper bound of the Bell inequality \cite{Ichikawa:2008eq}.   
Furthermore, the experiment \cite{CPLEAR:1998} by using the neutral $K$ meson pairs are not the Bell test, because it measures not $S^Q(\Delta t)$, but the correlation function $C^Q(t_1, t_2)$ where $|t_1-t_2|=\Delta t$.

In contrast, the ATLAS experiment admits independent measurements of the decay times, allowing for the selection process to completely close the locality loophole in the Bell test.  The remaining issue, from the viewpoint of standard Bell test, is the efficiency loophole, and at the moment we need to rely on the fair sampling assumption for it.  In this respect, improvement of detection efficiency is a desideratum, either through increase in the available decay modes or enhancement of the signal selection processing.

As a technical remark, in this simulation study, we have assumed $p_{\mathrm{T}}$ above 4~GeV, while the minimum $p_{\mathrm{T}}$ threshold of the di-muon trigger used in normal physics data-taking in the ATLAS Run~2 operation was 13~GeV.  Thus, the feasibility of reducing the threshold to $\sim4$~GeV in low-$\mu$ operation has to be studied further in the experiment. Even in case that $p_{\mathrm{T}}$ threshold has to be set higher, it would not be an issue for the measurement, if we take the data more than 1~fb$^{-1}$, which was assumed in Sec.~\ref{simulation}.

Despite these remaining issues, it seems almost certain that the ATLAS experiment offers a promising venue for the Bell test in the high energy domains with entanglement of much heavier particles,  and this will open up a new arena for experimental studies of quantum foundations hitherto unexplored.

\begin{acknowledgments}
We would like to thank Kazunori Hanagaki and Osamu Jinnouchi for useful discussions. This work was supported by JSPS KAKENHI Grant Number JP20H01906 and MEXT Quantum Leap Flagship Program (MEXT Q-LEAP) Grant Number JPMXS0120319794. 
\end{acknowledgments}

\appendix*

\section{Bell inequality with stochastic parameters without free will}
\label{q_cond_cor}

We here outline the argument that can still lead to the Bell inequality formally without using the analogy between the parameters 
$t_1, t_2$ and $a, b$ (for detail, see \cite{Ichikawa:2008eq}).  

Assume that $t_1, t_2$ are the parameters characterizing the state of the $B$ meson pair and independent of the hidden variables $\lambda$ but not determined by the free will of the experimenter.
For the sake of distinction and also for our convenience, below we employ the more standard notation $P(A,B\, | \, t_1, t_2)$ for the conditional probability distribution $P_{t_1, t_2}(A,B)$ used in the text.  

First, we note that, in the presence of the the hidden variables $\lambda$, 
the most basic element in our stochastic theory will be furnished by the conditional probability distribution $P(A,B \, | \, t_1, t_2,\lambda)$ together with $P(\lambda\,|\, t_1, t_2)$.  When combined, they provide the measurable distribution $P(A,B\, | \, t_1, t_2)$ by
\begin{equation}
P(A,B\, | \, t_1, t_2)=\int P(A,B \, | \, t_1, t_2,\lambda)P(\lambda\,|\, t_1, t_2)\, d\lambda.
\label{prob_t}
\end{equation}
To see this, one just recalls the multiplication law of the probability,
\begin{eqnarray}
&&\!\!\!\!\!\! P(A,B \, | \, t_1, t_2,\lambda)P(\lambda\,|\, t_1, t_2)\nonumber\\
&&=\frac{P(A,B, t_1, t_2,\lambda)}{P(t_1, t_2, \lambda)}\cdot\frac{P(t_1, t_2,\lambda)}{P(t_1, t_2)}\nonumber\\
&&=\frac{P(A,B, t_1, t_2,\lambda)}{P(t_1, t_2)}\nonumber\\
&&=P(A, B, \lambda \,|\, t_1, t_2)
\end{eqnarray}
to find
\begin{eqnarray}
&&\int P(A,B \, | \, t_1, t_2,\lambda)P(\lambda\,|\, t_1, t_2)\, d\lambda\nonumber\\
&&=\int P(A, B, \lambda \,|\, t_1, t_2)\, d\lambda\nonumber\\
&&=P(A, B \,|\, t_1, t_2),
\end{eqnarray}
which shows \eqref{prob_t}.

Next, the locality assumption leads to the following factorization:
\begin{equation}
P(A,B\, | \, t_1, t_2, \lambda)=P(A \, | \, t_1, t_2, \lambda)P(B \, | \, t_1, t_2, \lambda),
\end{equation}
meaning that the flavor of one of the two mesons at $t_1$ makes no influence on that of the other at $t_2$ and vice versa.
We are now required to impose stronger assumptions both on the independence of the decay times and locality, respectively. 
One of them is that the decay times $t_1, t_2$ have no correlation with the hidden variables:
\begin{equation}
P(\lambda\,|\, t_1, t_2)=P(\lambda).
\label{stat_ind}
\end{equation}
In other words, the decay times are statistically independent of the hidden variables.
This condition (\ref{stat_ind}) is called homogeneity condition in \cite{Ichikawa:2008eq}.
For the locality, we suppose that the decay time of one meson does not affect the flavor of the other (independence condition in \cite{Ichikawa:2008eq}): 
\begin{eqnarray}
P(A \, | \, t_1, t_2, \lambda)&=&P(A \, | \, t_1, \lambda),\nonumber\\
P(B \, | \, t_1, t_2, \lambda)&=&P(B \, | \, t_2, \lambda).
\end{eqnarray}

Then, we are allowed to proceed formally in a completely analogous manner as we did in the argument of the Bell inequality.  
In fact, we observe that the conditional correlation function reads
\begin{eqnarray}
C(t_1, t_2)&=&\sum_{A,B}AB\,P(A,B\, | \, t_1, t_2)
\nonumber\\
&=&\int A(t_1, \lambda)B(t_2, \lambda)P(\lambda)\, d\lambda,
\label{cor_condst}
\end{eqnarray}
where
\begin{eqnarray}
\!\!\!\!\!\! A(t_1, \lambda)&=&P(A=1\, | \, t_1, \lambda)-P(A=-1\, | \, t_1, \lambda),\nonumber\\
\!\!\!\!\!\! B(t_2, \lambda)&=&P(B=1\, | \, t_2, \lambda)-P(B=-1\, | \, t_2, \lambda).
\end{eqnarray}
From this we see immediately that the conditional correlation function $C(t_1, t_2)$ defined in \eqref{cor_condst} 
formally takes the same form as $C(a, b)$ in \eqref{lrtcor} under the identification of $t_1, t_2$ with $a, b$.  Consequently, 
we are able to arrive at the Bell inequality \eqref{Bell}, at least formally, without using the free will in the choice of the parameters.
This can be used to argue similarly that the violation of the formal Bell inequality implies the incompatibility of our assumptions with QM.
Obviously, the price we paid for this derivation lies in the additional assumptions required, weakening our statement considerably. 

\bibliography{reference}
  
\end{document}